# Status Report on Project GRAND: An Extensive Air Shower Array of Proportional Wire Chambers

J. Poirier, J. Carpenter, S. Desch, J. Gress, T. F. Lin, Y. Lu, and A. Roesch
*Physics Department, 225 NSH, University of Notre Dame, Notre Dame, IN 46556, USA*

## Abstract

Project GRAND is an extensive air shower array utilizing position sensitive detectors of proportional wire chambers. The 64 detectors deployed in a field 100 m x 100 m are located at 86.2° West, 41.7° North, at 220 m above sea level. The project was completed about two years ago and has been taking data, simultaneously, on two triggers: 1) multiple-hut coincidence triggers which collect data on extensive air showers, and 2) 200 Hz triggers which collect the single tracks stored in each station during the last 5 msec.

## 1 Introduction:

Project GRAND is located immediately north of the University of Notre Dame, 150 km east of Chicago, USA. The proximity to the University makes it ideal for graduate and undergraduate students to participate in the research. As well, the graphical nature of the data makes it conceptually easy for students to understand its character and facilitates their involvement in the physics of the experiment.

## 2 Experimental Array:

There are 64 detection stations arranged in an 8 by 8 grid in a field 100 m x 100 m (Fig. 1). The detectors, housed in huts 2.4 m wide, 2.4 m deep, by 1.5 m high (Fig. 2), are proportional wire chambers (PWCs). Each PWC contains two orthogonal x,y-planes. Each plane has an active area of 1.25 $m^2$. Each station is comprised of four x,y-paired PWC chambers arranged 200 mm above each other with a 50 mm steel plate placed just above the bottom x,y-pair (Linsley et al., 1987; Poirier et al., 1990). The steel plate allows the identification of muons; muon tracks penetrate the 50 mm steel (40 $g/cm^2$ = 2.9 radiation length) without scattering and without stopping 96% of the time if their energy is sufficient. This threshold energy for vertical muons is 0.1 GeV, increasing like $1/\cos(\theta)$ for $\theta$-angles deviating from vertical. In contrast, electrons stop 85% of the time and scatter ~11% of the time, thus leaving about 4% which mimic the pattern of a muon (muons also mimic electrons about 4% of the time) (Gress et al., 1990). Fig. 3 illustrates, from left to right: a stopping, scattering, and showering electron track and a noninteracting muon track on the right. Fig. 4 is a histogram of the number of stations in a shower trigger which require a > 5-hut coincidence.

The proportional wire chamber planes are 1.25 $m^2$ in area with four pairs of mutually orthogonal planes to a station. Each plane has 80 detection cells 14 mm wide. The cell size is ± 10 mm high and ± 7 mm wide. The eight planes of a station are read out in a serial fashion; each station thus has 640 bits of information. The 64 stations are read into a central data acquisition system in parallel at 10 MHz; the entire field of 41000 bits is read into storage memory in the DAS in < 70 microsec. Monte Carlo techniques were used in the design of the experiment; these calculations show that the angular resolution for an extensive air shower is ≤ 0.25°, and the average resolution for single muon tracks is ± 0.26° (Gress et al., 1991).

## 3 Construction and Operation:

The array is currently collecting data on two separate triggers at the same time: 1) shower data collected with a trigger demanding five stations in time coincidence; and 2) single track data collected every 5 millisec. These single tracks are stored in all the stations and are not discarded unless a new track overwrites it or the hut's information has been read out. The stations refresh their tracks at an average rate of 170 Hz. Monte Carlo studies indicate that: the first trigger collects shower data with primary energies

> 50 TeV; and the second trigger collects data on single muons which are produced by primary cosmic rays with primary energies between 10 and 300 GeV. Monte Carlo calculations (Fasso & Poirier, 1999) indicate that energies for single muon tracks come from primaries of energy between 10 and 300 GeV, an energy region above satellite data and below ground-based Cerenkov counters.

The presence of a muon in cosmic ray data is traditionally taken as a signature that the primary is *not* a gamma ray. However, a primary gamma ray of 100 GeV has about a 1.5% chance of yielding a muon at detection level, primarily through its small but non-negligible hadron production cross section. This probability rises as the energy increases (Fasso & Poirier, 1999). GRAND is thus able to detect gamma rays in this energy region through single tracks of identified muons and measure their angles.

Project GRAND was constructed over ~7 years. This long time span was largely due to the low rate at which construction funds became available. The total project cost was a million dollars which would have been less given a shorter construction period. The proportional wire chambers were manufactured with mass production techniques that achieved high tolerances, good uniformity, and economy. As a consequence, the PWC detectors have a wide voltage plateau allowing each quarter's detectors all to be operated at the same high voltage. Glass was chosen for the insulating material because it is a good insulator with uniform thickness and is economical. Signal wires are of unplated tungsten with rather tight specifications on radius and ovality. The resulting PWC operation is quite stable and uniform. All the PWC chambers in each quarter of the field operate from a single high voltage supply at a common voltage ~ 2600 volts at 25 microamps.

In the first seven years of operation, some of the difficulties were: 1) failure of a heater in a hut. In extreme cold temperature, the differential expansion of the signal wires (relative to the glass frame) can break the tungsten signal wires; 2) failure of the dehumidifiers, allowing the humidity to rise above 50% in the huts; and 3) a flood which destroyed 10 PWCs. A broken signal or high voltage wire in a PWC necessitated replacing the PWC. Thus during 1800 PWC-years of operation, 17 PWC chambers have been replaced. The flooding problem was solved by adding drains, leaving less than 0.4% failures per year. By solving two problems: 1) better tensioning of higher strength tungsten wire which would not break from differential shrinkage even at the coldest temperatures; and 2) a vapor barrier on the edge of the styrofoam preventing it from adsorbing the water vapor in humid air, these chambers could operate without heaters or dehumidifiers. Thus, the power necessary to operate a hut would be drastically reduced and could easily be accommodated with storage batteries and solar cells. This power could be further reduced by storing the hut's data and time information in local memory so they need not be transmitted to a central point.

The design, construction, and debugging of Project GRAND's data acquisition system completed partial requirements of two advanced degrees: a Ph.D. in Physics and a joint Master's degree in Physics and Electrical Engineering. Its design uses, mechanically, a VME backplane. There are four modules: 1) The coincidence unit receives signals from each of the 64 detectors, performs their coincidence, and attaches the time of the trigger to this data. 2) The memory unit receives and stores the status of each wire of the entire experiment (41000 bits) which is later transferred to the VME-based computer. 3) Upon a decision of the coincidence module, the address unit reads out the entire field of data by sending a 10 MHz train of clock signals for 70 microseconds to the 64 stations. This clock train is used to serially clock out the data from the shift register memory on the PWCs in the stations. Besides being triggered from the coincidence unit, it is also triggered on a regular basis at a rate of 200 Hz in order to collect the single tracks stored in the stations. 4) The WWVB clock interface performs two functions; one is the reading and storing of the WWVB time from a radio receiver tuned to WWVB in Boulder, Colorado precise to 1.0 millsec, absolute. The other is an internal 1 Mhz oscillator which is scaled. This latter function is a backup clock, which is quite precise although not absolute. Occasionally, the WWVB unit does not keep accurate time when, for example, it loses radio contact for too long a time. In this case the backup clock can be used and made absolute by comparing it to the accurate WWVB time preceding the incident.

An Omnibyte Module was selected for the monoboard computer to control the experiment online. It is based on a 68020 computer chip and operates on a VME bus. The program is run using its internal debugger code stored on a prom-chip. Basically, the program is read in through the RS232 port serially, the beginning address is supplied, and the computer asked to begin running at that address. Originally it was designed to accept each event of 5120 bytes and immediately write them to an 8 mm magnetic tape. Data runs began when four huts were completed; four huts were insufficient to study extensive air showers but running experience and some initial data were obtained by running the project on single-track triggers from any hut. As construction continued and huts were added, the data swamped the off-line computer which had to sift through all the bits of data to select the single tracks which had been collected. Finally, the offline computing was taking longer than the online computer was using to take the data.

Later a scheme was devised whereby the online CPU, which was faster than the offline CPU, searched for "one-and-only-one hit" in each of the eight planes of a given hut. Some 96% of data obtained with this online criterion later proved to be a single straight track in the offline analysis, which was now much faster with this pre-selection. This solved the offline CPU problems. When more huts were added, the online CPU restricted the data because it was not able to analyze the data as quickly as it was being obtained. Finally, eight additional on-line computer modules were added. These CPUs operated in a simple parallel-computing mode where each CPU in its turn handles the data from an event trigger and is told to search for "one-and-only-one hit" in each of the eight planes. Each CPU finds these muon candidates and stores the wire numbers of the hit wires in each plane. Muons are stored in a 900-muon buffer in each CPU. The eight CPUs in parallel are sufficient so each one has completed its computing task before it is asked to handle its next event. The online CPU deadtime was essentially eliminated using this parallel technique. The master CPU handles the shower data by itself (almost no computing is required and the rate is slow) as well as feeding the single-track trigger data to the respective node computers. It also asks which node CPUs are full and, when full, writes their data to an 8 mm magnetic tape. This master CPU writes data from the shower trigger to a separate 4 mm magnetic tape, thus separating the data from the two triggers from the very beginning of the analysis procedure.

## 4   Summary:

In summary, this array of position sensitive proportional wire chambers has proved to be economical to construct, requires little manpower to run, and the data are easy to analyze. Precision alignment of all planes yielding an absolute angle calibration was relatively easy and is constant. The detectors are reliable and require little maintenance. Gas tanks and data tapes need to be changed every two days. The angular precision for the primary cosmic ray is three times more precise than a comparable array built with plastic scintillators using timing techniques. Each detector, besides measuring angles, is also a muon detector, thus distributing muon detection uniformly throughout the array. Those tracks identified as muons are closer in angle to the primary and can be studied as single tracks. The array not only studies extensive showers > 50 TeV but also is simultaneously sensitive to showers in the 10 to 300 GeV range which yield single muon tracks at ground level. The references to Project GRAND's papers (prior to this conference) are listed after Fig. 4. References to papers in this conference: HE.3.1; HE.3.2.09, 10, 20; OG.2.3.12, OG.3.1.12, and OG.4.4.02. Gratitude is extended to all who have participated in Project GRAND's construction, operation, and analysis.

This research is presently being funded through grants from the University of Notre Dame and private donations. The National Science Foundation participated in GRAND's construction.

URL: www.nd.edu/~GRAND/          Phone: 219 631 7588
Email: GRAND@grand2.hep.nd.edu   Fax:   219 631 5952

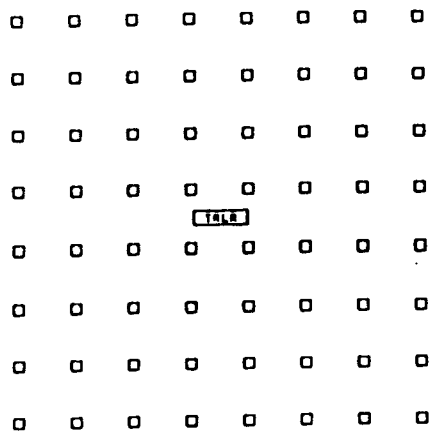

Figure 1: Layout of Project GRAND's 64 stations in 100 m x 100 m field.

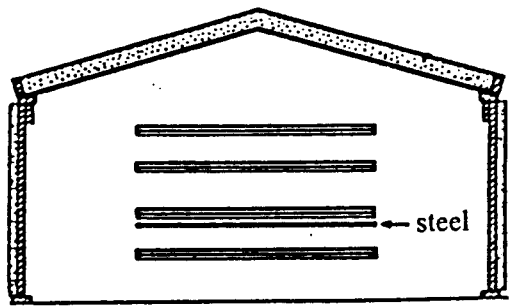

Figure 2: Vertical section of one station showing 4 PWCs (2 planes each) with 50 mm of steel above the bottom PWC.

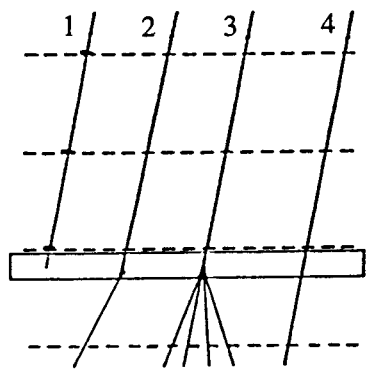

Figure 3: A vertical section showing the topology of electrons (1, 2, 3) interacting in steel and a noninteracting muon (4).

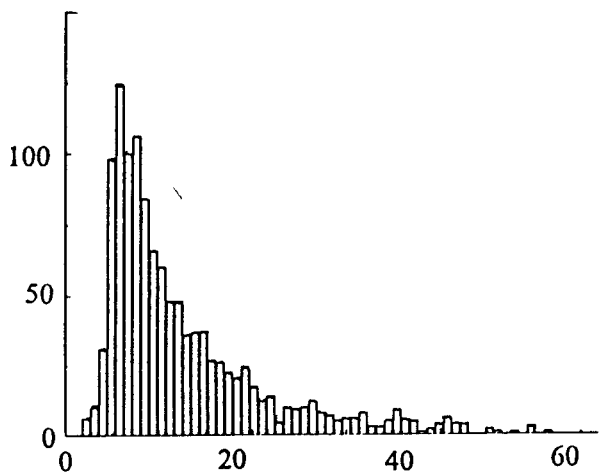

Figure 4: Histogram of the number of stations hit in a trigger. The low side cut-off is from the coincidence hardware.

## Project GRAND Publications

*Angle distributions* (MC): 20-ICRC 5, 452 (87); Jour Phy G 13, L217 (87); Phys Rev D36, 1378 (87); 21-ICRC 9, 46 (90); Jour Phy G 17, L19 (91)

*Composition:* 24-ICRC 3, 532 (95)

*Electronics:* Nucl Instr Meth A307, 425 (91); 22-ICRC 4, 421(91)

*General:* 20-ICRC 2, 438 (87); Nucl Instr Meth A260, 280 (87); Nucl Instr Meth A264, 81 (88); 21-ICRC 9, 1 (90); 21-ICRC 9, 50 (90); 21-ICRC 9, 126 (90); 21-ICRC 10, 335 (90); Nucl Phys B (Proc Suppl) 14A, 143 (90); Jour Phy G 17, 1303 (91); 22-ICRC 4, 417 (91); Jour Phy G 18, 1849 (92); A I P Conf Proc 276, 560 and 614 (93); 23-ICRC 4, 359 (93)

*M.C., timing arrays:* Nucl Instr Meth A257, 473 (87); Jour Phy G 13, 85 (87); 20-ICRC 6, 75 (87)

*Muons, identification:* Jour Phy G 13, L163 (87); Phys Rev D36, 1381 (87); A I P Conf Proc 220, 242 (91); Nucl Instr Meth A302, 368 (91); 22-ICRC 4, 425 (91)

*Results:* 22-ICRC 1, 192 (91); 22-ICRC 4, 291 (91); 22-ICRC 4, 295 (91); 22-ICRC 4, 583 (91); A I P Conf Proc 276, 566 (93); 23-ICRC 4, 422 (93); 23-ICRC 4, 454 (93); 24-ICRC 1, 634 (95); 24-ICRC 2, 144 (95); 25-ICRC 3, 289 (97); 25-ICRC 6, 221 (97); 25-ICRC 6, 225 (97); 25-ICRC 6, 365 (97); 25-ICRC 6, 369 (97)

*Authors:* Anagnostopoulos, Canough, Cunningham, Fields, Fischer, Funk, Gress, Herczeg, Kinney, Kochocki, Linsley, LoSecco, Lu, Markiewicz, Mikocki, Olson, Poirier, Rettig, Rohrer, Trzupek, and Wrotniak